\documentclass[twocolumn,secnumarabic,amssymb, aps, prl, superscriptaddress]{revtex4-2}

\usepackage{graphics, graphicx}
\usepackage[nomath]{stix}
\usepackage[italic]{mathastext}
\usepackage{bm}
\usepackage{mathtools,amsmath}
\usepackage{adjustbox}
\usepackage{array,booktabs,tabularx}
\usepackage{siunitx}
\usepackage{upquote}
\usepackage{hhline}
\usepackage{stackengine}
\usepackage{xcolor}
\usepackage{setspace} 
\usepackage{hyperref}
\usepackage{microtype}
\usepackage[capitalize]{cleveref}
\usepackage{xr}
\usepackage{float}
\usepackage{xspace}
\usepackage{xcolor}
\hypersetup{
    colorlinks,
    linkcolor={blue!50!black},
    citecolor={blue!50!black},
    urlcolor={blue!80!black}
}

\newcommand{\mue}{$\mu^{\rm e}$\xspace}
\newcommand{\muh}{$\mu^{\rm h}$\xspace}
\newcommand{\sigmace}{$\sigma_{\rm c}^{\rm e}$\xspace}
\newcommand{\sigmach}{$\sigma_{\rm c}^{\rm h}$\xspace}

\newcommand{\sigmash}{$\sigma_{\rm s}^{\rm h}$\xspace}
\newcommand{\xih}{$\xi^{\rm h}$\xspace}

\newcommand{\units}{$(\hbar/2e)e^2/h$\xspace}
\newcommand{\unitc}{$e^2/h$\xspace}
\newcommand{\ef}{E\textsubscript{F}\xspace}
\newcommand{\SCC}{Sc$_2$CCl$_2$\xspace}
\newcommand{\YCB}{Y$_2$CBr$_2$\xspace}

\newcommand{\frs}{F.R.S.\xspace}

\begin{document}

\begin{abstract}

  The conversion efficiency from charge current to spin current via spin Hall
    effect is evaluated by the spin Hall ratio (SHR).
  Through state-of-the-art \textit{ab initio} calculations involving both charge
    conductivity and spin Hall conductivity, we report the SHRs of the III-V
    monolayer family, revealing an ultrahigh ratio of 0.58 in the hole-doped GaAs
    monolayer.
  In order to find more promising 2D materials, a descriptor for high SHR is proposed and
    applied to a high-throughput database, which provides the fully-relativistic
    band structures and Wannier Hamiltonians of 216 exfoliable monolayer semiconductors and
    has been released to the community.
  Among potential candidates for high SHR, the MXene monolayer \SCC is identified
    with the proposed descriptor and confirmed by computation, demonstrating the descriptor validity for
    high SHR materials discovery.

\end{abstract}

\title{Enhanced Spin Hall Ratio in Two-Dimensional Semiconductors}

\author{Jiaqi Zhou} \email{jiaqi.zhou@uclouvain.be} 
\affiliation{Institute of Condensed Matter and Nanosciences (IMCN), Université catholique de Louvain (UCLouvain), 1348 Louvain-la-Neuve, Belgium}

\author{Samuel Poncé}
\email{samuel.ponce@uclouvain.be}  
\affiliation{Institute of Condensed Matter and Nanosciences (IMCN), Université catholique de Louvain (UCLouvain), 1348 Louvain-la-Neuve, Belgium}
  \affiliation{European Theoretical Spectroscopy Facility}
  \affiliation{WEL Research Institute, Avenue Pasteur 6, 1300 Wavre, Belgium}

\author{Jean-Christophe Charlier} \email{jean-christophe.charlier@uclouvain.be} 
\affiliation{Institute of Condensed Matter and Nanosciences (IMCN), Université catholique de Louvain (UCLouvain), 1348 Louvain-la-Neuve, Belgium}

\date{\today} \maketitle

\paragraph*{Introduction --}
The Hall effect encompasses a wide range of phenomena which realize the
  conversion between various physical properties~\cite{Inoue2005Sep,
    Chang2016Feb}.
The strength of Hall effect can be denoted by $\beta = \tan(\theta_{\rm H}) =
    {E_{\rm H}}/E$ where $\theta_{\rm H}$ is the Hall angle, $E_{\rm H}$ is the
  transverse Hall field, and $E$ is the longitudinal electric
  field~\cite{Ashcroft1976}.
Correspondingly, the strength of spin Hall effect (SHE) is given by the spin
  Hall ratio (SHR) as $\xi = \tan(\theta_{\rm{SH}}) = \frac{2e}{\hbar} \big|
    \frac{J_{\rm s} } {J_{\rm c} } \big|$ where $\theta_{\rm{SH}}$ is the spin Hall
  angle, $J_{\rm s}$ is the transverse spin Hall current density, and $J_{\rm c}$
  is the longitudinal charge current density.
SHR is often used as a proxy to indicate the charge-to-spin conversion
  efficiency which is crucial for low-power-consumption spintronic
  applications~\cite{Sinova2015Oct,Manchon2019Sep,Guo2021Jun}.
Indeed, when $\theta_{\rm{SH}}$ is small, the first-order Taylor polynomial
  gives $\xi \approx \theta_{\rm{SH}}$, which is a good approximation for the
  bulk semiconductors and metals where $\xi \sim
    0.01$~\cite{Zutic2004Apr,Olejnik2012Aug,Tao2018Jun}.
Recently, enhanced SHR has been found in van der
  Waals materials with strong spin-orbit coupling (SOC).
Huge SHRs over 10 are reported in topological
  insulators~\cite{Dc2018Sep,Khang2018Sep} while large SHR $\sim0.5$ in
  MoTe\textsubscript{2} and WTe\textsubscript{2} Weyl semimetals have also been
  theoretically and experimentally
  identified~\cite{Zhou2019Feb,Zhao2020Mar,Vila2021Dec,Song2020Mar}.
However, the relative abundance of topological materials is around 1~\%~\cite{Marrazzo2019Dec, Grassano2024Feb},
limiting material options for device manufacturing. 
The various two-dimensional (2D) materials enable a preferable compatibility with the integrated circuit~\cite{9581315}
with desirable properties.  For instance,
  the MoS\textsubscript{2} monolayer can exhibit $\xi = 0.14$ induced by
  the Rashba-Edelstein effect~\cite{Shao2016Dec}.
Note that large $\xi$ will break the approximation $\xi\approx
    \theta_{\rm{SH}}$ and therefore the spin Hall ratio rather than the spin Hall
  angle should be used to denote the ratio of spin current to charge current. 
2D materials composed of heavy atoms are promising for SHE~\cite{Wang2023Apr,Wang2023Aug}
since the strong SOC can induce a large spin Hall conductivity (SHC),
and doping is an effective way to manipulate the transport behaviors in semiconductors.
Both factors can promote the SHR enhancement in 2D semiconductors. 

Although charge transport and SHC have been separately investigated in 2D
  materials~\cite{Feng2012Oct,Sohier2018Nov,Wang2020Jan,Backman2022Dec,Zhang2023Feb},
  the study of SHR remains elusive due to the multidisciplinary complexity
  involving the electron-phonon interaction (EPI) for electron
  motion~\cite{Cepellotti2022Aug,Giustino2017Feb} and SOC for spin
  transport~\cite{Manchon2019Sep,Sinova2015Oct}.
In this Letter, we report the spin Hall ratios in monolayer semiconductors
  using density functional theory (DFT)~\cite{Kohn1999Oct}, density functional
  perturbation theory (DFPT)~\cite{Baroni2001Jul}, and Wannier
  functions~\cite{Marzari2012Oct}. 
The family of III-V monolayer semiconductors (MX, M=Ga, In, and X = As, Sb) are
  investigated.
In the hole-doped regimes, the charge conductivities are significantly
  suppressed by the strong inter-peak scattering, while high SHCs occur due to
  the strong SOC, yielding an ultrahigh SHR of $\xi=0.58$ in GaAs monolayer.
Taking the transport behaviors of III-V monolayers as a prototype, we
  propose a general descriptor for the high SHR based on the electronic
  structures. 
To validate the generality of this descriptor, we create a high-throughput
  database by performing fully-relativistic DFT calculations and Wannierizations
  on 216 monolayer semiconductors, whose electronic band structures, effective masses, and
  SHCs have been calculated.
The database is screened by the descriptor and suggests two MXene candidates,
  \SCC and \YCB monolayers.
The charge conductivities are investigated in both materials, confirming high
  SHR and validating the proposed SHR descriptor. 

\paragraph*{Methods --}
Charge transport properties are computed by solving the iterative Boltzmann
  transport equation~\cite{Ponce2020Feb} and the spin Hall conductivity using the
  Kubo formula~\cite{Qiao2018Dec} as implemented in the \textsc{Quantum
    ESPRESSO}~\cite{Giannozzi_2017}, \textsc{EPW}~\cite{Ponce2016Dec,Lee2023Feb},
  and \textsc{Wannier90}~\cite{Pizzi_2020} codes considering SOC and 2D Coulomb
  truncation~\cite{Sohier2017Aug}.
The high-throughput calculations are implemented using
  \text{AiiDA}~\cite{Huber2020Sep}, \textsc{Pseudo Dojo}~\cite{vanSetten2018May},
  and the \textsc{MC2D} database~\cite{MC2D1,MC2D2}.
Additional details are provided in Secs.~\ref*{supp-sec:method} and
  \ref*{supp-sec:HT} of the Supplemental Material (SM)~\cite{SI2023}. 
For the III-V and MXene monolayers investigated in this work, details of the relaxed
  atomic structures, effective masses, densities of states (DOS), doping levels, and
  electron and phonon dispersions are given in SM~\cite{SI2023}.
The input and output files, pseudopotentials, as well as Wannier Hamiltonians
  and band structures of 216 monolayer semiconductors, are provided on Materials
  Cloud Archive~\cite{MCA}.

\begin{figure}
  \includegraphics[width=0.35\paperwidth]{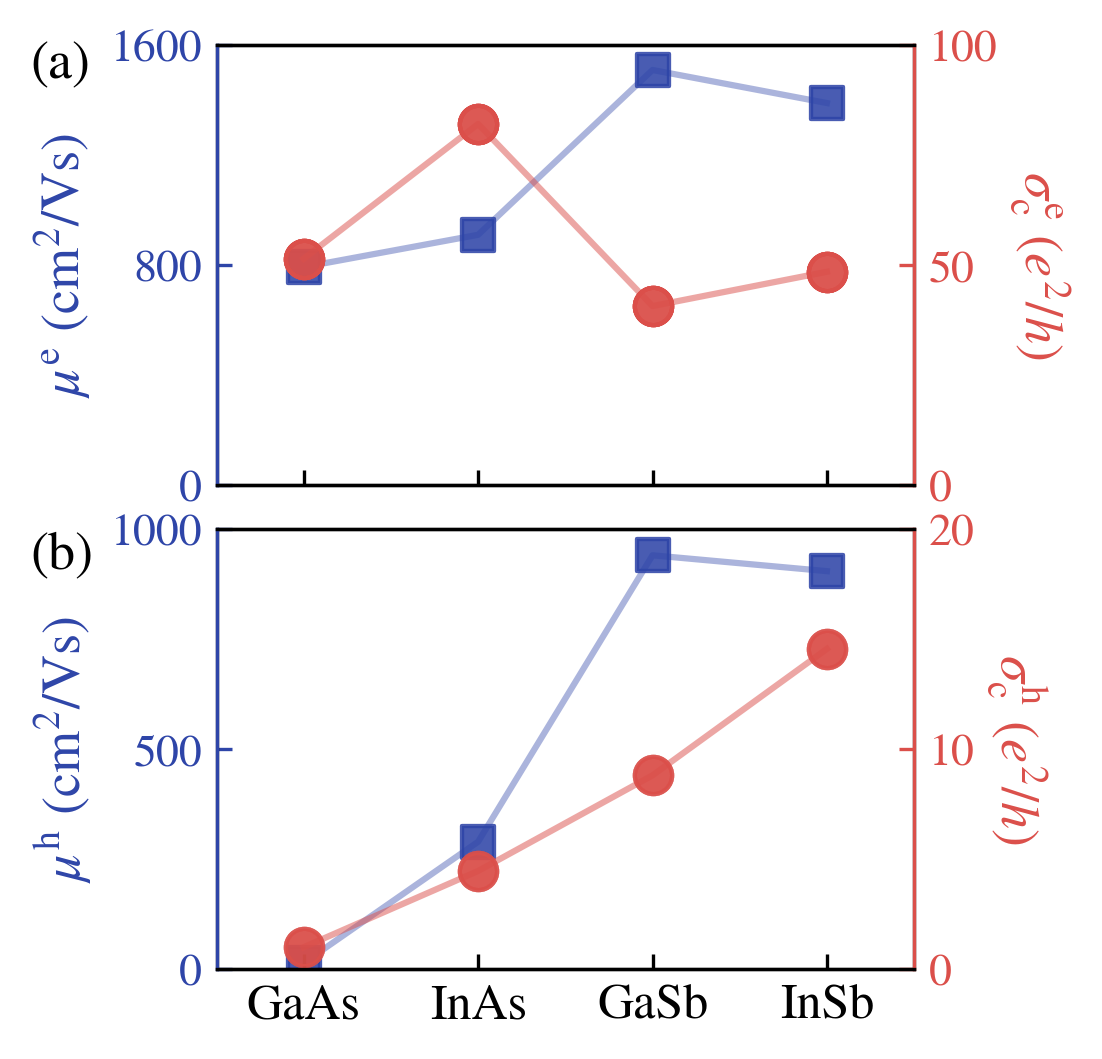}
  \caption{\label{fig:Fig1}
    Drift mobilities of pristine semiconductors (blue) and charge conductivities of
      doped systems (red) at 300~K.
    \mue~and~\muh denote (a)~electron and (b)~hole mobilities of
    pristine semiconductors with  square markers  (left~axis),
    \sigmace~and~\sigmach indicate the charge conductivities of
    (a)~electron-doped  and (b)~hole-doped systems with  circle markers  (right~axis).
  }
\end{figure}

\paragraph*{Charge transport --}
The phonon-limited charge conductivity in doped 2D semiconductor is calculated
  as~\cite{Ponce2020Feb} \begin{align}\label{eq:sigma} \sigma_{\alpha\beta} =
    \frac{-e}{S^{\mathrm{uc}}} \sum_n \int \frac{\mathrm{d}^2
      \mathbf{k}}{\Omega^{\mathrm{BZ}}} v_{n\mathbf{k}\alpha} \partial_{E_{\beta}}
    f_{n\mathbf{k}},\end{align} where $\alpha$ and $\beta$ are Cartesian
  directions, $S^{\rm uc}$ is the unit cell area, $\Omega^{\rm BZ}$ is the first
  Brillouin zone area, and $v_{n\mathbf{k}\alpha} = \hbar^{-1} \partial
    \varepsilon_{n\mathbf{k}}/\partial k_{\alpha}$ is the band velocity, $n$ is the
  band index.
The linear variation of the electronic occupation function $f_{n\mathbf{k}}$ in
  response to $\mathbf{E}$, $\partial_{E_{\beta}}f_{n\mathbf{k}} $, can be
  obtained by solving the Boltzmann transport equation [\cref*{supp-eq:bte} of
      SM~\cite{SI2023}] which induces the scattering rate given by
  \begin{multline}\label{eq:scattering_rate} \tau_{n\mathbf{k}}^{-1} =
  \frac{2\pi}{\hbar} \sum_{m\nu} \!
\int\! \frac{\mathrm{d}^2 \mathbf{q}}{\Omega^{\text{BZ}}} | g_{mn\nu}(\mathbf{k,q})|^2 \\ \times \big[ (n_{\mathbf{q}\nu} +1 - f_{m\mathbf{k+q}}^0) \delta( \varepsilon_{n\mathbf{k}} - \varepsilon_{m\mathbf{k+q}}   -  \hbar \omega_{\mathbf{q}\nu})\\
  + (n_{\mathbf{q}\nu}  + f_{m\mathbf{k+q}}^0 )\delta(\varepsilon_{n\mathbf{k}} - \varepsilon_{m\mathbf{k+q}} + \hbar \omega_{\mathbf{q}\nu}) \big],
\end{multline}
where $g_{mn\nu}(\mathbf{k,q})$ is the electron-phonon matrix element with  phonon frequency $\omega_{\mathbf{q}\nu}$,
$\varepsilon_{n\mathbf{k}}$ and $ \varepsilon_{m\mathbf{k+q}}$ are eigenvalues,
$f_{n\mathbf{k}}$ is the Fermi-Dirac distribution,
$n_{\mathbf{q}\nu}$ is the Bose-Einstein distribution.
The drift mobility of pristine semiconductor is related to the charge
  conductivity as $\mu_{\alpha\beta} = \sigma_{\alpha\beta} / (en^{\rm c}) $ when
  the carrier density $n^{\rm c}$ is very small such that ionized impurity
  scattering can be neglected. 
Due to crystal symmetry, $\mu = \mu_{xx} = \mu_{yy}$, $\sigma = \sigma_{xx} =
    \sigma_{yy}$ in all the III-V monolayers.
Note that $\mu$ and $\sigma$ are separately calculated since in this work, heavy dopings
are applied, thus breaking the linear relation between them~\cite{Ma2014Mar}.
In the following, $\sigma_{\rm c}$ is used to denote the charge conductivity.

\Cref{fig:Fig1} presents the room-temperature mobilities of the pristine monolayers
and the conductivities of doped systems.
All the materials exhibit high electron mobilities, thanks to the small
  electron effective mass and the single valley in the conduction bands. 
More variations in the hole mobility are observed due to the multi-peak band
  structures. 
The two arsenides present different mobilities of 14 and 289, while two similar
  values, 940 and 904, are observed in the antimonides.
A detailed analysis of mobility mechanisms is given in our accompanying
  manuscript~\cite{PRB2024}.
In this paper,  the conductivities in doped III-V monolayers are extensively explained.

Doping is a practical method to tune the transport properties of
  semiconductors~\cite{Awate2023Mar}.
Sufficient carriers are induced by heavy doping which turns semiconductors into
  metallic systems where SHE can occur.
Considering the DOS, an electron doping of $1\times
    10^{13}$~cm$^{-2}$ and a hole doping of $2\times 10^{13}$~cm$^{-2}$ are
  respectively applied to the III-V monolayers, whose structures have been
  relaxed again.
The main impact of such doping is the shift of Fermi energy
  (E\textsubscript{F}) by a few hundred meV, leaving the crystal structure and
  electronic bands nearly unaffected as verified by \cref*{supp-sec:struct} of SM~\cite{SI2023},
  where the phonon dispersions are also presented. 
The absence of imaginary frequency in phonon dispersions demonstrates the
  stabilities of doped systems.
Interestingly, an electron-hole asymmetry occurs in the phonon dispersions of
  all the III-V monolayers: the phonon dispersion is weakly affected by electron
  doping, while softening specifically occurs in the hole doping case.
The asymmetry can be explained by the difference in conduction and valence
  bands.
For the electron doping, the $\Gamma$ valley is occupied, leading to a moderate
  screening of the EPI by free carriers and a decrease of the EPI.
In contrast, multiple inequivalent peaks are occupied at room temperature in
  the case of hole doping.
In this regime, the EPI is \emph{strengthened} with doping due to the charge
  transfer between inequivalent peaks with opposite deformation
  potentials~\cite{Sohier2019Aug,Sohier2023Jan,CPB2019}, resulting in the observed phonon
  softening in the optical modes. 
In a nutshell, the asymmetry in the conduction and valence bands leads to the
  reduced EPI in electron-doped systems  as verified by the high \sigmace in
      \cref{fig:Fig1}(a)  and the enhanced EPI in hole-doped systems as verified by  the low
      \sigmach in \cref{fig:Fig1}(b).

\begin{figure}[t]
  \includegraphics[width=0.4\paperwidth]{ 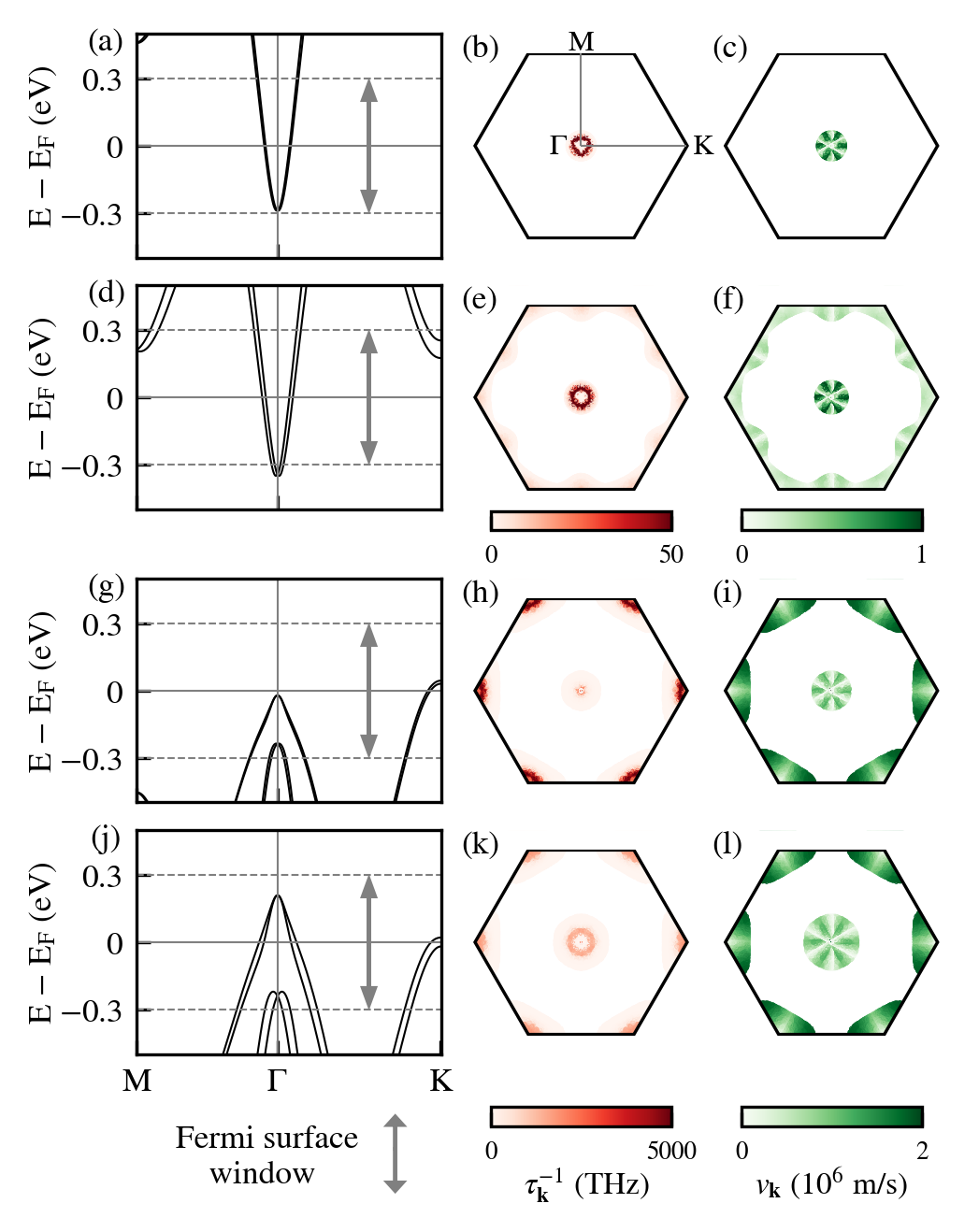}
  \caption{\label{fig:Fig2}
    Electronic structures, $\mathbf{k}$-resolved scattering rates, and
      $\mathbf{k}$-resolved velocities for charge conductivities of
      (a)-(c)~electron-doped GaAs, (d)-(f)~electron-doped GaSb, (g)-(i)~hole-doped
      GaAs, and (j)-(l)~hole-doped GaSb.
    The vertical arrow  denotes the Fermi surface window, where
    0.3~eV has been validated for the  transport property convergence. 
  }
\end{figure}

It is expected that the charge conductivity should be proportional to the
  carrier mobility.
However, \cref{fig:Fig1}(a) illustrates that GaAs presents a much lower \mue
  but higher \sigmace than GaSb with the same electron doping.
In \cref{fig:Fig1}(b), GaSb shows a high \muh but a low \sigmach with the hole
  doping.
These unusual behaviors will be interpreted within the self-energy relaxation
  time approximation~\cite{Ponce2020Feb}, where the conductivity is inversely
  proportional to the scattering rate and directly proportional to carrier
  velocity.
Considering the Fermi-Dirac distribution at equilibrium and 300~K, we define
  the $\mathbf{k}$-resolved scattering rates as $ \tau^{-1}_{\bf{k}} = \sum_n
    \frac{-\partial f_0}{\partial \varepsilon_{n \bf{k}}} {\tau^{-1}_{n \bf{k}}}$,
  and $\mathbf{k}$-resolved velocities as $ v_{\bf{k}} = \sum_n {v_{n \bf{k}}}$,
  where $n$ denotes the number of bands involved in the transport.
\Cref{fig:Fig2} compares the electronic structures,   $ \tau^{-1}_{\bf{k}}$,  and $ v_{\bf{k}} $ of doped GaAs and GaSb.
Before doping, both pristine GaAs and GaSb exhibit a Rashba splitting~\cite{Wu2021Jan} 
in the conduction band minimum (CBM) which can be regarded as a single valley. 
After electron doping, \cref{fig:Fig2}(a) shows that for GaAs, the single valley
  is preserved in the Fermi surface window [\ef$-~0.3$~eV, \ef$+~0.3$~eV], while
  \cref{fig:Fig2}(d) shows that for GaSb, more states around M and K points enter
  into the window, leading to enhanced scatterings with states that possess low
  velocities.
As a result, \sigmace in GaSb is reduced as shown in \cref{fig:Fig1}(a).
The surprising behaviors of \muh and \sigmach of GaSb in \cref{fig:Fig1}(b) can
  be attributed to the doping-induced \ef shift. 
The valence band maximum (VBM) of pristine GaAs locates at K points, leading to a
  multi-peak band structure and a high DOS around VBM, thus the hole doping can
  only induce a small \ef shift as shown in \cref{fig:Fig2}(g).
Besides, the spin-orbit splitting at K in the electronic band is 12~meV, which
  matches well with the phonon energy at K in the phonon dispersion.
Considering momentum and energy conservations, the strong inter-peak
  scatterings between the K and K' peaks are allowed in both pristine and
  hole-doped GaAs.
Differently, the VBM in pristine GaSb is located at $\Gamma$, which is a single
  peak for the hole mobility and leads to a high \muh.
The single-peak band structure gives a small DOS around VBM, thus a large \ef
  shift of 0.21~eV is induced by the hole doping as shown in \cref{fig:Fig2}(j),
  leading the states around K to dominate the scattering.
Considering the low velocities at K points, \sigmach is greatly reduced
  compared with \muh in GaSb.
It should be noted that for GaSb, the spin-orbit splitting at K is 40~meV,
  mismatching the phonon energy, thus the inter-peak scattering between K and K'
  is weakened compared with GaAs, as shown by the colors in
  Figs.~\ref{fig:Fig2}(h) and (k).
The discussions above demonstrate that doping is an effective way to manipulate
  the electronic structure, further controlling the EPI and charge conductivity
  in semiconductors.

\begin{figure}[t]
  \includegraphics[width=0.4\paperwidth]{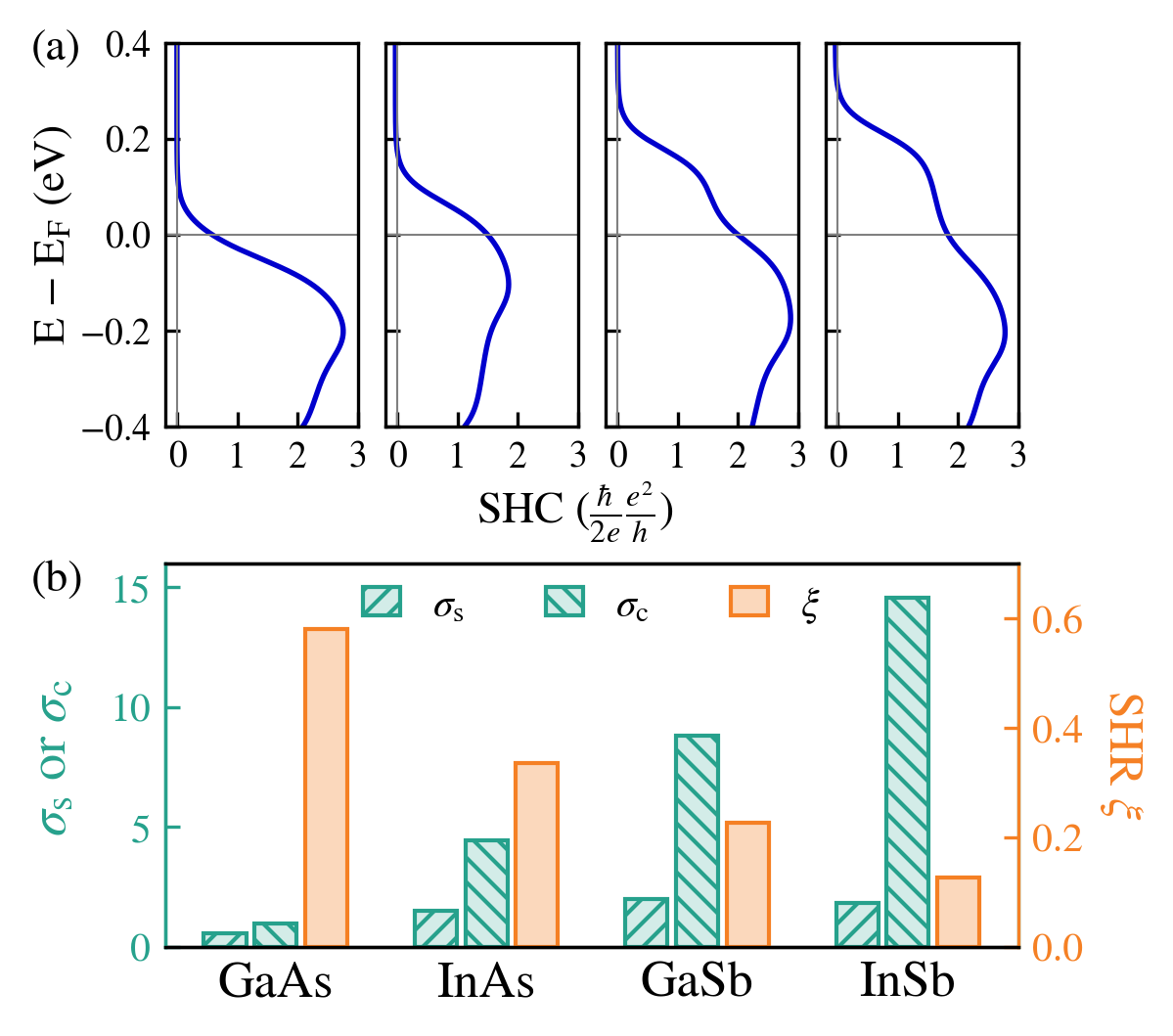}
  \caption{ \label{fig:Fig3}
    (a)~Energy-dependent SHCs at 300~K.
    (b)~SHCs  $\sigma_{\rm s}$ at \ef (unit: \units),  
    charge conductivities $\sigma_{\rm c}$ (unit: \unitc),
       and SHRs $\xi$ of hole-doped
    GaAs, InAs, GaSb, and InSb,
    respectively.
  }
\end{figure}

\paragraph*{Spin Hall conductivity --}

SHE is composed of intrinsic and extrinsic contributions.
The former is defined by the Kubo formula, and the latter is due to the
  skew-scattering and side-jump mechanisms, strongly dependent on
  disorder~\cite{Lowitzer2011Feb,Gorini2015Aug,Shashank2023Feb}.
In this work, we focus on the intrinsic SHC in a weak scattering limit.
With spin current along $x$, electric field along $y$, and spin orientation
  along $z$ direction, the intrinsic SHC is calculated using Kubo
  formula~\cite{Qiao2018Dec}: \begin{equation} \label{eq:kubo_shc} \sigma_{\rm s}
    = \frac{\hbar}{2e} \frac{e^2}{\hbar} \int_\text{BZ} \frac{\mathrm{d}^2
      \mathbf{k}}{(2\pi)^2} \Omega_{\rm s}(\mathbf{k} ), \end{equation} where
  $\Omega_{\rm s}(\mathbf{k} )= \sum_{n} f_{n\mathbf{k}}\Omega_{{\rm
          s},n}(\mathbf{k}) $ is the spin Berry curvature (SBC) with $ f_{n\mathbf{k}}$
  at 300~K and the band-resolved SBC as \begin{equation} \label{eq:spinberry}
    \Omega_{{\rm s},n}(\mathbf{k}) = {\hbar}^2 \sum_{m \ne n}
    \frac{-2\operatorname{Im}[\langle n\mathbf{k}| \hat{j}_z |m\mathbf{k}\rangle
        \langle m\mathbf{k}|
        \hat{v}_{y}|n\mathbf{k}\rangle]}{(\varepsilon_{n\mathbf{k}}-\varepsilon_{m\mathbf{k}})^2
      + \eta^2}, \end{equation} where $\hat{j}_z = \frac{1}{2} \{\hat{\sigma}_z
    \hat{v}_x + \hat{v}_x \hat{\sigma}_z \}$ is the spin current operator,
  $\hat{\sigma}_z $ is the Pauli operator, $ \hat{v}_x$ and $ \hat{v}_y$ are
  velocity operators.
Considering the SOC strength in all the materials, a broadening of $\eta =
    2$~meV is used as a weak scattering in spin transport~\cite{Li2015Apr}. 
Due to the low conductivities, the hole-doped systems are more promising for
  high SHRs, thus III-V monolayer with $2\times 10^{13}$~cm$^{-2}$ hole doping are
  targeted for SHCs, and this doping level has been realized in 2D
  systems~\cite{Shi2023May,Sohier2019Aug}.
\Cref{fig:Fig3}(a) presents the energy-dependent SHCs of hole-doped materials at room temperature. 
GaAs exhibits $\sigma_{\rm s}^{\rm h} = 0.6 $~\units due to the small \ef shift
  induced by doping.
More prominently, SHCs of InAs, GaSb, and InSb can reach up to 1.5, 2.0, and
  1.8~\units, respectively. 
For comparison, MoS$_2$ monolayer only reaches $\sigma_{\rm s}^{\rm h} \approx
    0.2$~\units~\cite{Feng2012Oct}.
The high SHCs in hole-doped III-V monolayers are attributed to the in-gap \ef
  location, as shown in \cref{fig:Fig2}(j).
The spin-orbit gap separates the positive and negative SBC.
When the \ef locates inside the spin-orbit gap, SHC, as the integration over
  the Fermi sea of SBC, can be maximized by the sign-invariant SBC over all the
  $\mathbf{k}$-points. 
The SBC decompositions of all the materials are presented in
  \cref*{supp-sec:SBC} of SM~\cite{SI2023}.
For GaAs, the SBC originates from both K and $\Gamma$ points, while for the
  others, the SBC mainly stems from the $\Gamma$ point due to the in-gap \ef at
  $\Gamma$.
The discussions above highlight that the doping in semiconductors can yield
  large SHCs in III-V monolayers.

\begin{figure}[tb]
  \includegraphics[width=0.42\paperwidth]{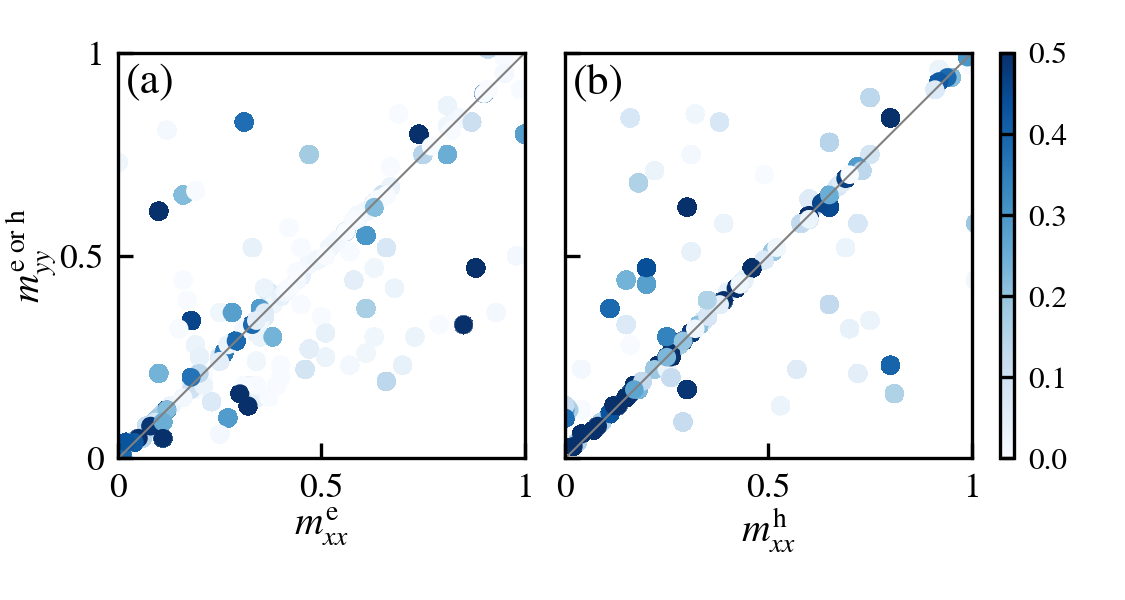}
  \caption{\label{fig:Fig4} 
    Electron (a) and hole (b) effective masses of 216 monolayer semiconductors
    in unit of $m_0$, SHC values are given by the color bar in units of \units
    in electron- and hole-doped systems, respectively. 
    Diagonal lines denote isotropic materials where $m_{xx} = m_{yy}$.
    The doping concentration is $2\times 10^{13}$~cm$^{-2}$ for both carriers.  
  }
\end{figure}

\paragraph*{Spin Hall ratio --}
With the charge conductivities $\sigma_{\rm c}$ and spin Hall conductivities
  $\sigma_{\rm s}$, the spin Hall ratio $ \xi = \frac{2e}{\hbar} \big|
    \frac{\sigma_{\rm s}}{\sigma_{\rm c}} \big|$ can be discussed in hole-doped
  systems.
\Cref{fig:Fig3}(b) shows that large \sigmash values are found in antimonides which also possess
fairly large \sigmach, as a result, GaSb and InSb 
exhibit \xih = $0.23$ and $0.13$, respectively.
More importantly, hole-doped arsenides are perfect candidates with high SHCs
  and low charge transports, yielding exceptional $\xi^\mathrm{h} = 0.58$ and
  $0.34$ in GaAs and InAs, respectively.
Compared with heavy metals where $\xi\approx0.01$, the hole-doped GaAs and
  other III-V monolayers exhibit great potential for efficient charge-to-spin
  conversion.

\paragraph*{Spin Hall ratio descriptor --}
Taking the transport behaviors in III-V monolayers as a prototype, a descriptor
  to enhance SHR can be proposed.
The idea is to decrease the charge conductivity as well as increasing SHC.
Large effective mass reduces $\sigma_{\rm c}$ via lowering down the carrier
  velocity.
As discussed in \cref{fig:Fig2}, the band structure with multiple extrema
  around \ef can enhance scattering, and the matching between energy difference
  $\Delta \varepsilon_{n{\rm K}}$ and phonon energy $\omega_{n{\rm K}}$ is a plus
  to strengthen the inter-peak scattering, note that K can be any $\bf k$-point
  away from $\Gamma$ in the BZ.
Regarding the SHC, the energy-dependent SHCs have shown the importance of the
  \ef location.
Overall, a descriptor for high SHR can be proposed: large effective mass,
multiple inequivalent extrema in the electronic band structure, and
a \ef located inside the spin-orbit gap. 
It should be stressed that within a limited doping concentration, there is a
  competition between the first two conditions and the last one, since flat and
  multiple band extrema would induce a large DOS, which hinders the tuning of
  \ef.
Consequently, a delicate balance between $\sigma_{\rm c}$ and $\sigma_{\rm s}$
  is essential.

\begin{figure}[tb]
  \includegraphics[width=0.4\paperwidth]{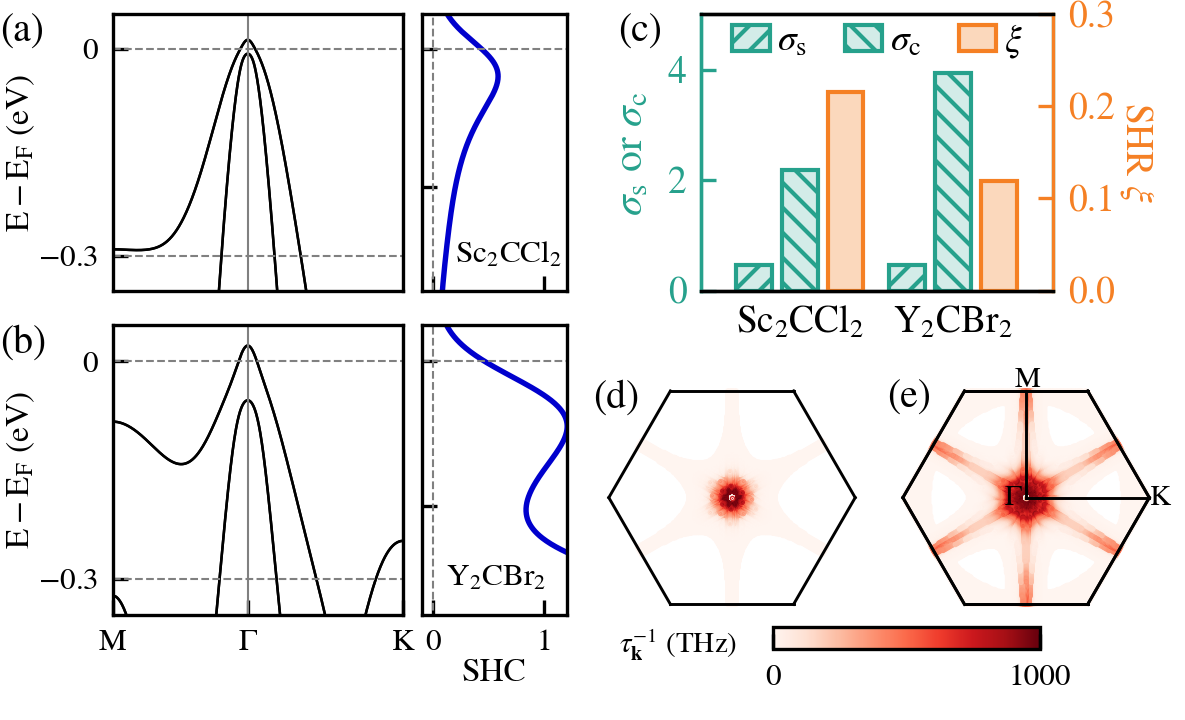} \caption{\label{fig:Fig5}
  Hole-doped \SCC and \YCB monolayers:
  (a)-(b)~Electronic structures and energy-dependent SHCs in units of \units.
  (c)~Spin Hall and charge conductivities as well as  SHRs.
  (d)-(e)~Scattering rates   $ \tau^{-1}_{\bf{k}}  $.
  }
\end{figure}

\paragraph*{High-throughput calculations --}
Potential candidates for high SHR can be found by screening a database with the aforementioned
  descriptor.
Focusing on the 2D materials, we perform fully-relativistic high-throughput
  calculations for exfoliable monolayers.
From the \textsc{MC2D} database~\cite{MC2D1,MC2D2}, all the rare-earth free
  materials with up to 6 atoms per unit cell are considered, yielding 216
  semiconductors which are employed for the fully-relativistic DFT calculations
  and Wannierizations. 
\Cref{fig:Fig4}  shows the electron and hole
effective masses of all the semiconductors,
as well as  SHC values  in the doped systems.
The effective mass is calculated by $ m_{\alpha\alpha} =({\frac{1}{\hbar^2}
        \frac{d^2 \varepsilon}{{dk^2_\alpha}}})^{-1}$ ($\alpha = x, y$) through the
  interpolation around CBM and VBM respectively for electron and hole. 
Overall, the hole presents larger effective mass than the electron, and
  hole-doped materials exhibit larger SHC values.
Nevertheless, many materials with high SHCs present small effective masses,
  populated at the left lower corner in \cref{fig:Fig4}(b).
Still, by setting the screening conditions as $ m_{xx} \approx m_{yy} >
    0.5~m_0$, multiple band extrema around \ef, and SHC > 0.4 \units, two MXene
  monolayers are selected as candidates, Sc$_2$CCl$_2$ and Y$_2$CBr$_2$.
Quality assessment of Wannier Hamiltonians, effective masses, doped SHCs,
  energy gaps, and band structures of all the 216 semiconductors are reported in
  \cref*{supp-sec:HT} of SM~\cite{SI2023}.
Notably, 97\% materials present the interpolation error of less than 10~meV for bands 
below CBM~+~1~eV, demonstrating the high accuracy of the database. 

\SCC and \YCB present   similarities
since they are composed by elements from the same group.
Both are easily exfoliable with a small interlayer binding energy~\cite{MC2D2}.
After a full relaxation including the hole doping, both present a
  \text{P-3m1} space group, and their mechanical stabilities have been validated
  by phonon dispersion without imaginary frequency as given in
  \cref*{supp-fig:BANDS_SCC} of SM~\cite{SI2023}.
Figures~\ref{fig:Fig5}(a)-(b) illustrate the hole-doped band structures of \SCC and
  \YCB.
$ m_{xx} = m_{yy} \approx 0.7~m_0$ on the top band.
Heavy elements lead to the SOC splitting and two peaks around the $\Gamma$
  point. 
\ef locates inside the gap, leading to \sigmash=~0.5~\units.
Interestingly, the band structures can be modeled by the Luttinger Hamiltonian,
  which shows robust SHC against disorder~\cite{Murakami2004Jun}.  
Benefiting from all the advantages above, the SHR in \SCC is computed to be
  \xih = 0.22, and \xih = 0.12 for \YCB as shown in \cref{fig:Fig5}(c).
The discrepancy in SHR values is attributed to the different charge transports.
Figures~\ref{fig:Fig5}(d)-(e) show that the scattering in \SCC reaches up to
  836~THz thanks to its small SOC gap of 20~meV matching phonon energy, while the
  scattering in \YCB is limited to 370~THz due to its large gap of 75~meV. 
As a result, \SCC presents lower \sigmach and higher SHR.
The discussions above validate the SHR descriptor and reveal \SCC as a
  potential candidate for high SHR.
By expanding the high-throughput calculations database, materials with more
  ideal electronic structures could be found to further enhance SHR.

\paragraph*{Experimental feasibility --}
Charge-to-spin conversion has been realized in MoS$_2$ and WSe$_2$ monolayers
  grown by chemical vapor deposition~\cite{Shao2016Dec}.
Since the first synthesis of 2D AlN layers by metal organic
  deposition~\cite{Wang2019Jan}, many efforts have been devoted to the synthesis
  of other 2D III-V materials~\cite{Lu2022MariScience, Nikolaevich2023May}.
For example, GaSb films can be grown via a seeded lateral epitaxy, and the
  free-standing crystalline GaSb can be exfoliated from these
  films~\cite{Manzo2022Jul}.
Moreover, 2D InAs flakes with high crystalline quality have been synthesized
  through van der Waals epitaxy with a thickness down to
  4.8~nm~\cite{Dai2022Nov}.
Due to chemical similarity in one family, we expect similar techniques can be
  applied to the other III-V monolayers.
Two promising materials \SCC and \YCB are easily exfoliable  from their van der Waals bulk compounds~\cite{MC2D2}.
Finally, the doping levels proposed in this work can be realized via the
  advanced technique of electron beam, which implements the doping of $1.7\times
    10^{13}$~cm$^{-2}$ in 2D systems~\cite{Shi2023May}.
The doped state persists even after removing the electron beam and back-gate
  voltage, and the process is reversible and repeatable~\cite{Shi2023May}.
Moreover, the doping level over $5\times 10^{13}$~cm$^{-2}$ has been realized
  in MoS$_2$ monolayers via the ionic-liquid gate~\cite{Sohier2019Aug}.

In conclusion, 2D materials with high spin Hall ratios have been found using a
  multidisciplinary investigation involving charge transport, spin Hall
  conductivity, and high-throughput database.
The hole-doped GaAs monolayer presents an ultrahigh SHR of $\xi = 0.58$,
  attributed to the strong scattering and the high SHC.
A SHR descriptor is proposed and validated by a high-throughput database of 216
  exfoliable monolayer semiconductors, suggesting a new promising material \SCC.
Besides, this database is fully released to the community. 
This work reveals potential 2D materials for efficient charge-to-spin
  conversion, providing a guideline for materials discovery in spintronics.

% \textit{Acknowledgements --}
\begin{acknowledgments}
The authors would like to thank Xi Dai, Matteo Giantomassi, and Junfeng Qiao for fruitful discussions.
S.~P.
acknowledges the support from the Fonds de la Recherche Scientifique de Belgique (\frs-FNRS).
J.~Z. and J.-C.C. acknowledge financial support from the F\'ed\'eration Wallonie-Bruxelles through the ARC Grant ``DREAMS'' (No. 21/26-116), from the EOS project ``CONNECT'' (No. 40007563), and from the Belgium \frs-FNRS through the research project (No.~T.029.22F).
Computational resources have been provided by the PRACE award granting access
  to MareNostrum4 at Barcelona Supercomputing Center (BSC), Spain and Discoverer
  in SofiaTech, Bulgaria (OptoSpin project ID. 2020225411), and by the Consortium des Équipements de Calcul Intensif (C\'ECI), funded by the \frs-FNRS under Grant No. 2.5020.11 and by the Walloon Region, as well as computational resources awarded on the Belgian share of the EuroHPC LUMI supercomputer.
\end{acknowledgments}


%\bibliographystyle{apsrev4-2-author-truncate}
%\bibliography{ref.bib}
\end{document}